\documentstyle[11pt,mrs2003,epsfig]{article}
\begin{document} 
\def \lta {\mathrel{\vcenter
     {\hbox{$<$}\nointerlineskip\hbox{$\sim$}}}}
\def \gta {\mathrel{\vcenter
     {\hbox{$>$}\nointerlineskip\hbox{$\sim$}}}}

\title{QUINTESSENCE AND THE DARK MATTER ABUNDANCE}

\author{FRANCESCA ROSATI}
\affil{Dipartimento di Fisica and INFN, Universit\'a di Padova, via Marzolo 8, 35131 Padova, Italy } 
 
\begin{abstract} 
Results of a recent study \cite{paper} on how the Quintessence scalar could affect the relic abundance of dark matter particles are presented. 
\end{abstract}

\section{Introduction}

The standard cosmological scenario  assumes that before Big Bang Nucleosynthesis 
(BBN), the universe was dominated by radiation with the 
Hubble parameter evolving like $H^2 \sim \rho_r \sim a^{-4}$, 
where $\rho_r$ is the energy density of radiation and $a$ is the scale 
factor of the universe.
However, if we imagine to add a significant fraction of scalar 
energy density $\rho_\phi$ to the background radiation at some time in the past, 
and if $w_\phi > w_r=1/3$, then $\rho_\phi$  
would decay more rapidly than $\rho_r$, but temporarily increase the global
expansion rate. A measurable effect of this modification to the standard scenario 
is the anticipation of the 'freeze-out' time of neutralinos and the consequent 
enhancement of their relic abundance, as explicitly calculated in \cite{salati}.

In a flat universe, a scalar field with potential $V(\phi)$ obeys the 
equations
\begin{equation}
\ddot{\phi} + 3H\dot{\phi} + \frac{dV}{d\phi} = 0  \,\,\,  ;  \,\,\,\,\,\,\,\
H^2 \equiv \left( \frac{\dot{a}}{a} \right)^2 = \frac{8\pi}{3M_p^2}~ \rho  \,\,\, .
\label{scaleq}
\end{equation}
For any given time during the cosmological evolution,
the relative importance of the scalar energy density
w.r.t.~to matter and radiation in the total energy density 
$\rho \equiv \rho_{m} + \rho_{r} + \rho_{\phi}$
depends on the initial conditions, and is constrained by the available
cosmological data on the expansion rate and large scale structure.
If the potential $V(\phi)$ is of the runaway type, the initial stage of the scalar evolution is typically 
characterized by a period of so--called `kination' \cite{quint}
during which the scalar energy density
$\rho_{\phi} \equiv \dot{\phi}^2/2 + V(\phi)$
is dominated by the kinetic contribution $E_k= \dot{\phi}^2/2 \gg V(\phi)$,
giving $w_\phi =1$.
After this initial phase, the field comes to a stop and remains nearly 
constant for some time (`freezing' phase), until it eventually reaches an 
attractor solution \cite{quint}.

Then, if we modify the standard picture 
according to which only radiation plays a role in the post-inflationary era and 
suppose that at some time $\hat{t}$  the scalar contribution was 
small but non negligible w.r.t.~radiation, then at that time the 
expansion rate $H(\hat{t})$ should be correspondingly modified.
During the kination phase the scalar
to radiation energy density ratio evolves like 
$\rho_{\phi}/\rho_r \sim a^{-3(w_\phi - w_r)} = a^{-2}$, and so
the scalar contribution would rapidly fall off and leave room to 
radiation.
In this way, we can respect the BBN bounds and at the same time keep a 
significant scalar contribution to the total energy density just few red-shifts 
before\footnote{For example, a scalar to radiation ratio $\rho_{\phi}/\rho_r=0.01$ at BBN 
($z\simeq 10^9$) would imply $\rho_{\phi}/\rho_r =0.1$ at 
$z \simeq 3.16 \times 10^{9}$ and $\rho_{\phi}/\rho_r =1$ at $z\simeq 10^{10}$,
if the scalar field is undergoing a kination phase.}.
The increase in the expansion rate $H$ due to the additional scalar 
contribution would anticipate the decoupling of particle species and 
result in a net increase of the corresponding relic densities.
As shown in \cite{salati}, a scalar to radiation energy density ratio
$\rho_{\phi}/\rho_r \simeq 0.01$ at BBN would give an enhancement
of the neutralino codensity of roughly three orders of magnitude.

\section{The model}

The enhancement of the relic density of neutralinos requires that at some 
early time the scalar energy density was dominating the Universe. 
This fact raises a problem if we want to identify the scalar
contribution responsible for this phenomenon with the
Quintessence field \cite{paper}.  Indeed, the initial conditions
must be such that the scalar energy density is sub-dominant 
at the beginning, if we want the Quintessence field to reach the cosmological
attractor in time to be responsible for the presently
observed acceleration of the expansion \cite{quint}.
For initial conditions $\rho_{\phi}\gta \rho_r$ we obtain instead an
`overshooting' behaviour: the scalar field rapidly rolls down the potential
and after the kination stage remains frozen at an energy density much smaller than the critical one.
However, as shown in \cite{mpr}, more complicated dynamics are
possible if we relax the hypothesis of considering a single uncoupled scalar.  
The presence of several scalars and/or of a
small coupling with the dark matter fields could modify the
dynamics in such a way that the attractor is reached in time even if
we started, for example, in the overshooting region.

\noindent
{\bf More fields.}
Consider a potential of the form
$V(\phi_1,\phi_2) = M^{n+4} \left(\phi_1 \phi_2\right)^{-n/2}$,
with M a constant of dimension mass.
In this case, as discussed first in \cite{mpr}, the two fields' dynamics 
enlarges the range of possible initial conditions
for obtaining a quintessential behaviour today.
This is due to the fact that the presence of more scalars allows to play with 
the initial conditions in the fields' values, while maintaining the total 
initial scalar energy density fixed. 
Doing so, it is possible to obtain a situation in which for a fixed 
$\rho_{\phi}^{in}$ in the overshooting region, if we keep initially
$\phi_1=\phi_2$ we actually produce an overshooting behaviour, 
while if we choose to start with $\phi_1\not =\phi_2$ (and {\it the same} 
$\rho_{\phi}^{in}$) it is possible to reach
the attractor in time.
This different behaviour emerges from the fact that, if at the beginning
$\phi_2 \ll \phi_1$  then, in the example at hand,
$\partial V/\partial\phi_2 \gg \partial V/\partial\phi_1 $ and so 
$\phi_2$ (the smaller field) will run away more rapidly and tend to overshoot 
the attractor, while 
$\phi_1$ (the larger field) will move more slowly, join the attractor 
trajectory well before 
the present epoch and drive the total scalar energy density towards the 
required value.

\noindent
{\bf Interaction.}
Suppose now that the Quintessence scalar is not completely decoupled from the 
rest of the Universe.
Among the possible interactions, two interesting 
cases are the following:
\begin{equation}
V_b = \frac{b}{2}\ H^2 \phi^2 \;\;\;\;\;
\mbox{\rm or} \;\;\;\;\;
V_c = c \rho_m \phi
\label{inter}
\end{equation}
If we add $V_b$ or $V_c$ to $V=M^{n+4}\phi^{-n}$, the potential will acquire a (time-dependent) minimum and the scalar field will be prevented from running freely to infinity.
In this way, the long freezing phase that characterises the evolution of a 
scalar field with initial conditions in the overshooting region 
can be avoided.
Effective interaction terms like $V_b$ in eq.~(\ref{inter}) are discussed
 in Refs.~\cite{dine}, where it is shown that supersymmetry 
breaking effects in the early universe can induce mass corrections to the 
scalar Lagrangian of order $H^2$. 
The second type of interaction ($V_c$ in eq.~(\ref{inter})) can emerge in the
context of scalar-tensor theories of gravity (see, for example Ref.~\cite{damour}), in which a metric coupling
exists between matter fields and massless scalars.
The scalar field equation in this context is 
modified  by the presence of an additional source term
\begin{equation}
\ddot{\phi} + 3H\dot{\phi} + \frac{1}{2}\frac{dV}{d\phi} = 
-4\pi G \alpha(\phi) T
\label{eq-bd} 
\end{equation}
where $\alpha(\phi)$ is a generic function of $\phi$ and $T$ is the trace of the 
matter energy-momentum tensor $T^{\mu\nu}$. The case
$\alpha(\phi)=0$ corresponds to a scalar field 
decoupled from the rest of the world.

\vskip0.2cm
A more detailed discussion, together with numerical examples, can be found in Ref.~\cite{paper}.


\vfill 

\begin{thebibliography}{}

\bibitem{paper} F.Rosati, 2003, Phys. Lett. B {\bf 570}, 5.

\bibitem{salati} P.~Salati, 2002, astro-ph/0207396.

\bibitem{quint}
P.~J.~Steinhardt, L.~M.~Wang and I.~Zlatev, 1999,
Phys.\ Rev.\ D {\bf 59}, 123504;
S.~C.~Ng, N.~J.~Nunes and F.~Rosati, 2001, Phys.\ Rev.\ D {\bf 64}, 083510; 
A. de la Macorra and G. Piccinelli, 2000, Phys.\ Rev.\ D {\bf 61}, 123503.

\bibitem{mpr} A.~Masiero, M.~Pietroni and F.~Rosati, 2000,
Phys.\ Rev.\ D {\bf 61}, 023504.

\bibitem{dine}
M.~Dine, L.~Randall and S.~Thomas, 1995,
Phys.\ Rev.\ Lett.\ {\bf 75}, 398;
D.~J.~H.~Chung, L.~L.~Everett and A.~Riotto, 2003,
Phys.\ Lett.\ B {\bf 556}, 61.

\bibitem{damour}
T.~Damour, 1996, gr-qc/9606079.

\end{thebibliography}
\end{document}